\RequirePackage{silence}
\documentclass[online]{aa}    

\usepackage[varg]{txfonts}
\usepackage[dvipsnames]{xcolor}

\usepackage{graphicx}

\usepackage{hyperref}
\usepackage{hyperxmp}
\hypersetup{
    colorlinks=true,
    allcolors=blue,
    unicode=true,
    pdftitle={Sunspot simulations with MURaM - I. Parameter study using potential field initial conditions},
    pdfauthor={Markus Schmassmann, Nazaret Bello González, Rolf Schlichenmaier, Jan Jurčák},
    pdfsubject={Astronomy \& Astrophysics, 708, A123 (2026); DOI: 10.1051/0004-6361/202558785},
    pdfkeywords={Sun: photosphere, Sun: magnetic fields, sunspots, magnetohydrodynamics (MHD), methods: numerical},
    pdfpublication={Astronomy \& Astrophysics},
    pdfvolumenum={708},
    pdfpagerange={A123},
    pdfdoi={10.1051/0004-6361/20260000},
    pdfcontactemail={schmassmann@leibniz-kis.de},
}

\IfFileExists{latexml.sty}{\usepackage{latexml}}{\newif\iflatexml\latexmlfalse}
\newcommand{\hereOrThere}[2]{\iflatexml\providecommand{#1}{#2}\else\providecommand*{#1}{}#2\fi}
\iflatexml\else\fi

\usepackage{etoolbox}

\NewCommandCopy{\AALogoOld}{\AALogo}
\renewcommand*{\AALogo}{\href{https://www.aanda.org}{\AALogoOld}}

\makeatletter
\ifaa@online\idline{\ifnum\value{page}=1{\myFullArticleNumber\hspace{3.5cm}\textcolor{red}{arXiv version, typsetting different from A\&A}}\else{\aa@authorrunning: \myFullArticleNumber}\fi}\else\fi
\renewcommand*\aa@copyrightname{\copyright~The Authors}
\renewcommand\aa@manuscriptname{%
  manuscript no. \aa@numarticle
  \hspace{\stretch{1}}%
  \aa@copyrightname\ \the\year
}
\renewcommand\aa@textidlineempty{{\slshape A\&A proofs:}\ manuscript no.~\aa@numarticle}
\renewcommand*\doi[1]{%
  \renewcommand*\aa@doi{\href{https://doi.org/#1}{https://doi.org/#1}}%
  \renewcommand*\aa@doifig{#1}%
}
\ifaa@online\iflatexml\else%
    \fancypagestyle{firstpage}{%
        \fancyhf{}%
        \renewcommand*{\headrulewidth}{\z@}%
        \renewcommand*{\footrulewidth}{\z@}%
        \fancyfoot[RO]{\aa@footfont \aa@numarticle\aa@pageof}%
        \fancyfoot[LE]{\aa@footfont \aa@numarticle\aa@pageof}%
        \fancyfoot[LO]{\\[0.5\baselineskip]\includegraphics[height=16pt]{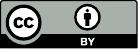}}
        \fancyfoot[C]{\\[0.5\baselineskip]\scalebox{0.78}{%
                \begin{tabular}{@{}c@{}}%
                    \tiny Open Access article, published on \href{https://www.arxiv.org}{arXiv}, under the terms of the Creative Commons Attribution License (\href{https://creativecommons.org/licenses/by/4.0}{https://creativecommons.org/licenses/by/4.0}),\\
                    \tiny which permits unrestricted use, distribution, and reproduction in any medium, provided the original work is properly cited.
                \end{tabular}\qquad%
            }%
        }%
    }%
    \fancypagestyle{otherpage}{%
        \fancyhf{}%
        \fancyhead[CO]{\aa@headfont \ifx\aa@idline\empty\aa@textidlineempty\else\aa@idline\fi}%
        \fancyhead[CE]{\aa@headfont \ifx\aa@idline\empty\aa@textidlineempty\else\aa@idline\fi}%
        \fancyfoot[RO]{\aa@footfont \aa@numarticle\aa@pageof}%
        \fancyfoot[LE]{\aa@footfont \aa@numarticle\aa@pageof}%
        \renewcommand*{\headrulewidth}{\z@}%
        \renewcommand*{\footrulewidth}{\z@}%
    }
    \pretocmd{\aa@maketitle}{\vspace*{0.3cm}}{}{}
\fi\else\fi
\newcommand{\latexMLheader}{\myFullArticleNumber\hfill\href{https://www.aanda.org}{Astronomy \& Astrophysics}\\\aa@doi\\\copyright~The Authors \the\year}
\makeatother

\newcommand*{\sethyphenpenalty}[1]{\hyphenpenalty=#1}
\doi{10.1051/0004-6361/202558785}
\AANum{A123} 
\newcommand*\myFullArticleNumber{A\&A, 708, A123 (2026)}
\yearCop={2026}
\authorlink={Markus Schmassmann, Nazaret Bello Gonzales, Rolf Schlichenmaier, Jan Jurcak}

\AddToHook{shipout/lastpage}{\label{LastPage}}
\begin{document}
\makeatletter
\renewcommand*\aa@pageof{, page \thepage{} of \pageref{LastPage}} 
\makeatother
\makeatletter\ifaa@online\iflatexml\latexMLheader\else\fi\else\fi\makeatother

   \title{Sunspot simulations with MURaM}
   \subtitle{I. Parameter study using potential field initial conditions}


   \author{Markus Schmassmann\inst{1}$^,$\thanks{Corresponding author: \href{mailto:schmassmann@leibniz-kis.de}{\texttt{schmassmann@leibniz-kis.de}}}\href{https://orcid.org/0000-0002-7303-1006}{\includegraphics[height=8pt]{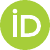}},
        Nazaret Bello González\inst{1}\href{https://orcid.org/0000-0002-0479-9134}{\includegraphics[height=8pt]{ORCID-iD_icon_vector.pdf}}, 
        Rolf Schlichenmaier\inst{1}\href{https://orcid.org/0000-0002-9220-4515}{\includegraphics[height=8pt]{ORCID-iD_icon_vector.pdf}}, and
        Jan Jurčák\inst{2}\href{https://orcid.org/0000-0002-7386-8578}{\includegraphics[height=8pt]{ORCID-iD_icon_vector.pdf}}}

   \authorrunning{M. Schmassmann, N. Bello González, R. Schlichenmaier, J. Jurčák}

   \institute{Institut für Sonnenphysik (KIS), Georges-Köhler-Allee 401a, 79110 Freiburg, Germany
        \and
        Astronomical Institute of the Czech Academy of Sciences, Fričova 298, 25165 Ondřejov, Czech Republic}

   \date{Received 24 December 2025 / Accepted 10 February 2026 }

 
  \abstract
   {Existing sunspot simulations fail to reproduce the observed magnetic field distribution due to an artificially increased $B_\mathrm{hor}$ at the upper boundary.}
   {We explore alternative ways to better reproduce the magnetic and dynamic properties of observed sunspots.}
   {We used the MURaM radiative magnetohydrodynamic code. As the initial conditions, we placed a potential magnetic field into small-scale dynamo simulations and used potential field extrapolation as the top boundary conditions.}
   {We find that 
   (1) simulations with increasing initial magnetic field strengths (20\,kG, 40\,kG, 80\,kG, and 160\,kG) show increasing spot, umbral, and penumbral sizes; 
   (2) penumbral-to-spot sizes are smaller than those measured in observed sunspots;
   (3) none of the runs show pure Evershed (radially outwards) flows and, instead, bi-directional flows with inflows in the inner penumbra and outflows in the outer penumbra were measured, consistent with observations of early stages of penumbra formation for runs with 80\,kG or more and 96/32\,km resolution, whereas runs with $\leq40\,$kG showed pure inflows;
   (4) simulations with 160\,kG and an increased resolution of 32/16\,km contain filaments with bi-directional and Evershed flows;
   (5) simulations with fluxes $>10^{22}$\,Mx show unrealistically strong fields in the umbra; and 
   (6) best runs with 160\,kG and $10^{22}$\,Mx give realistic profiles of $B_\mathrm{z}$ and $B_\mathrm{r}$ with radius, albeit with stronger fields than those typically observed. Finally, (7) increasing the width of the box and reducing the overall flux by subtracting a uniform opposing vertical field have little influence on internal spot dynamics and fields. Nonetheless, these choices affect the average vertical field beyond the spot.}
   {Simulations of small (10$^{22}$\,Mx) sunspots with an initial potential field and intensified magnetic field strength at the bottom of the box seem to best reproduce observational results of the initial stages of sunspot formation. Our findings also suggest that increased numerical resolution could be critical for achieving fully developed penumbrae.}
   
   \keywords{  Sun: photosphere --
               Sun: magnetic fields --
               sunspots --
               magnetohydrodynamics (MHD) --
               methods: numerical
            }
   \authorrunning{Schmassmann, M., et al.}
   \titlerunning{Sunspot simulations with MURaM: I}
   {\maketitle
   \iflatexml\else\nolinenumbers\setlength{\topsep}{0pt}\setlength{\itemsep}{0pt plus 1pt}\fi 
%
\section{Introduction} 
\label{sec:intro}
The art of magnetohydrodynamic (MHD) sunspot simulations has only
recently been developed. In 2007, \citet{Heinemann:2007} presented the first slab simulation of a sunspot, followed by similar simulations by \citet{Scharmer:2008} and \citet{2009ApJ...691..640R}. These simulations provided the first insights into the penumbral fine structure. Following a series of initial trials, particularly in \citep{2009Sci...325..171R}, \citet{2012ApJ...750...62R} managed to run MHD simulations of a circular sunspot that qualitatively reproduced the observed properties of penumbral fine structure. In other words, the structure of the penumbral filaments is comparable to the observational analysis of \citep{Tiwari:2013}.

However, the formation of an extended penumbra has been achieved in these MHD simulations by imposing artificial top boundary conditions that make the magnetic field more horizontal at the photospheric level. As shown by \citep{2020A&A...638A..28J}, the global magnetic properties of the resulting sunspot are not fully realistic. The magnetic field inclination on the umbral-penumbral boundary is around 60$^\circ$, which is significantly more horizontal than in the observed sunspots, where the highest values for the inclination are around 45$^\circ$ and occur only in large sunspots with more flux than the simulated sunspot.
Furthermore, the vertical magnetic field component, $B_\mathrm{z}$, at the umbral boundary in these simulations is below the value expected from observations, a $B_\mathrm{z}^\textrm{stable}$ of around 1.8~kG.
In \citet{2020A&A...638A..28J}, it was also shown that the inclination distribution as a function of the fractional spot radius, as well as $B_\mathrm{z}$ at the umbral boundary of observed sunspots, is comparable to an MHD simulation of a sunspot with a potential field as an initial condition. 
The use of a potential field as an initial condition for sunspot simulations was first proposed by \cite{Nordlund2015},  using slab geometry. Based on this idea, we performed an MHD simulation of a circular sunspot, as described by \citet{2020A&A...638A..28J}. To find the best agreement between the simulated and observed sunspots, we conducted a parameter study of sunspot simulations using potential field initial conditions.

\hereOrThere{\myTabSims}{\begin{table*}[htb]
    \caption{Simulation parameters.}
    \label{tab:sims}
    \centering
\newlength{\minusLength}\settowidth{\minusLength}{$-$}
\begin{tabular}{l|rrcccccc}\hline\hline\rule{0pt}{2ex}%
Sunspot run & $B_0$ & $B_\mathrm{opp}$ & $F_\mathrm{Gauss}$ & $F_\mathrm{tot}$ & Box width & $d_\mathrm{h}$ & Run time & $t_\mathrm{ref}$\\
& [kG] & [G] & [$10^{22}$Mx] & \hspace*{0.5\minusLength}[$10^{22}$Mx]\hspace*{0.5\minusLength} & [Mm] & [km] & [h] & [h]\\
\hline\rule{0pt}{2ex}%
\hphantom{1}20\_opp000o & 20.0 & 0 & 1 & 1       & 49.152 & 96 & 7.2038 & $2.8847-4.0839$\\
\hphantom{1}40\_opp000o & 40.0 & 0 & 1 & 1       & 49.152 & 96 & 7.1966 & $2.8815-4.0789$\\
\hphantom{1}80\_opp000o & 80.0 & 0 & 1 & 1       & 49.152 & 96 & 7.2167 & $2.8889-4.0911$\\
160\_opp000o & 160.0 &   0 & 1        & 1        & 49.152 & 96 & 14.429 & $2.8886-4.0908$\\
160\_opp000ho& 160.0 &   0 & 1        & 1        & 49.152 & 32 & 7.7131 & $7.2163-7.7131$\\
160\_opp000h & 160.0 &   0 & 1        & 1        & 49.152 & 32 & 7.9997 & $3.0149-4.0149$\\
160\_opp050c & 160.0 &  50 & 1.120796 & 1        & 49.152 & 96 & 4.0205 & $3.0142-4.0205$\\
160\_opp100c & 160.0 & 100 & 1.241592 & 1        & 49.152 & 96 & 4.0075 & $3.0075-4.0075$\\
160\_opp150c & 160.0 & 150 & 1.362388 & 1        & 49.152 & 96 & 4.0076 & $3.0076-4.0076$\\
160\_opp200c & 160.0 & 200 & 1.483184 & 1        & 49.152 & 96 & 4.0057 & $3.0057-4.0057$\\
160\_opp300c & 160.0 & 300 & 1.724776 & 1        & 49.152 & 96 & 4.0045 & $3.0045-4.0045$\\
160\_opp000  & 160.0 &   0 & 1        & 1        & 49.152 & 96 & 40.025 & $3.0111-4.0147$\\
160\_opp050  & 160.0 &  50 & 1        & 0.879204 & 49.152 & 96 & 4.0226 & $3.0182-4.0226$\\
160\_opp100  & 160.0 & 100 & 1        & 0.758408 & 49.152 & 96 & 4.0367 & $3.0167-4.0367$\\
160\_opp150  & 160.0 & 150 & 1        & 0.637612 & 49.152 & 96 & 4.0135 & $3.0123-4.0135$\\
160\_opp200  & 160.0 & 200 & 1        & 0.516816 & 49.152 & 96 & 4.0400 & $3.0223-4.0400$\\
160\_opp300  & 160.0 & 300 & 1        & 0.275224 & 49.152 & 96 & 4.0313 & $3.0198-4.0313$\\
160\_opp100w & 160.0 & 100 & 1        & 0.033632 & 98.304 & 96 & 4.0267 & $3.0199-4.0267$\\
160\_opp150w & 160.0 & 150 & 1        &\hspace*{-\minusLength}%
                                     $-$0.449552 & 98.304 & 96 & 4.0156 & $3.0156-4.0156$\\
160\_opp200w & 160.0 & 200 & 1        &\hspace*{-\minusLength}%
                                     $-$0.932736 & 98.304 & 96 & 4.0214 & $3.0138-4.0214$\\
160\_opp000b & 160.0 &   0 & 1        & 1        & 49.152 & 96 & 6.1536 & $2.9848-4.0010$\\\hline\rule{0pt}{2ex}%
alpha1.0     &   6.4 &     &          & 1.198126 & 49.152 & 32 & 6.5096 & $6.0096-6.5096$\\
alpha1.5     &   6.4 &     &          & 1.198126 & 49.152 & 32 & 6.4120 & $5.9120-6.4120$\\
alpha2.0     &   6.4 &     &          & 1.198126 & 49.152 & 32 & 6.4133 & $5.9133-6.4133$\\
alpha2.5     &   6.4 &     &          & 1.198126 & 49.152 & 32 & 6.4222 & $5.9222-6.4222$\\\hline\rule{0pt}{2ex}%
NOAA AR 11591&       &     &          &          &        & 362.5 & & 2012.10.18 10:00:00\\\hline%
\end{tabular}\vspace*{3mm}
\end{table*}}

The sunspot simulations of \citet{2021ApJ...907..102P} also adopted potential top boundary conditions. However, the penumbra of their sunspots was wide only between the nearby opposite \mbox{polarity} spots and very narrow in the perpendicular direction. Furthermore, their simulations are, by design, strongly affected by fluting instability, which makes them short-lived. The same authors also created simulations of starspots on other main \mbox{sequence} stars \citep{2020ApJ...893..113P}. 
\citet{2025A&A...693A.264B} simulated starspots, extending the work of \citet{2020ApJ...893..113P}, but again forcing the magnetic field at the top to be inclined, resulting in too horizontal fields, at least when comparing the G2V starspot to sunspot observations.

Hideyuki Hotta investigated the effect of higher resolution on sunspot simulations with a potential-field top boundary condition. Matthias Rempel joined this investigation and also \mbox{analysed} what happens when the resolution is changed during the simulation runs. 
All simulations in their investigation use \mbox{potential} field top boundary conditions.
While presentations of preliminary analysis \citep{2025ghh..confE..13R} are very encouraging, these results have not yet been published in a peer-reviewed journal. Furthermore, the high resolution used in these simulations results in prohibitively high costs.

In solar observations, the penumbra forms around exist\-ing flux concentrations \citep[e.g.][]{2010A&A...512L...1S}. Various authors have done flux emergence simulations 
\citep[e.g.][]{2010ApJ...720..233C,2017ApJ...846..149C,2020MNRAS.494.2523H}. 
These simulations show, consistently with solar surface observations, that deeply anchored bundles rise coherently up to around 2\,Mm below the surface, to then decay in small parts that rise with individual convective cells. At the surface, they \mbox{re-coalesce} to larger structures that end up forming sunspots \citep[see the ‘tethered-balloon’ model of][]{1981phss.conf...98S}.
However, they \mbox{either} manage to create a wide enough penumbra at most in some directions or have incorrect flow directions (inflows along the filaments and downflows at the umbral boundary) in most or all penumbral filaments.

In the following, Sect.~\ref{sec:simu} describes the simulation setups, \mbox{including} the initial conditions, their variables, and the values they take, as well as the data processing methods. Section~\ref{sec:results} presents the results, focusing on intensities, surface flows, and magnetic field properties. Section~\ref{sec:conclusions} provides a summary and conclusion, including a discussion of how the results relate to observations.

\section{Simulation set-ups and processing}
\label{sec:simu}

Our simulations were run using the MPS/University of Chicago Radiative Magneto-hydrodynamics code \citep[MURaM; ][]{2004A&A...421..741V,2009ApJ...691..640R}. 
We used the updated numerical treatment from \citet{2017ApJ...834...10R}, without using the coronal features introduced therein.
The bottom boundary is open to mass flows and the magnetic fields are extrapolated symmetrically, as described by \citet[][Sect. 2.2, item `OSb']{2014ApJ...789..132R}. The top boundary is open to inflows and closed to outflows, with the magnetic field potential as in \citet[][Appendix C.1]{2006PhDT.......338C} and \citet[][Appendix B]{2012ApJ...750...62R}. 

\hereOrThere{\myTabRes}{\begin{table*}[htb]
    \caption{Simulation results.}
    \label{tab:res}
    \centering
\newlength\tmpSep
\setlength{\tmpSep}{\tabcolsep}
\setlength{\tabcolsep}{2.525pt}
\begin{tabular}{l|rrrrr|rrr|rrrrrr}\hline\hline\rule{0pt}{2ex}%
Sunspot run & $r_\mathrm{u}$ & $r_\mathrm{s}$ & 
$r_\mathrm{pu}$/$r_\mathrm{s}$ & $F_\mathrm{spot}$ & $\mathrm{d}_tF/F$ & 
$\max B_\mathrm{z}$ & $\max B_\mathrm{r}$ & $r_{\max B_\mathrm{r}}$ &
$\min v_\mathrm{z}$ & $r_{\min v_\mathrm{z}}$ & 
$\min v_\mathrm{r}$ & $r_{\min v_\mathrm{r}}$ & 
$\max v_\mathrm{r}$ & $r_{\max v_\mathrm{r}}$\\
& [Mm] & [Mm] & &  \hspace*{-4pt}[$10^{22}$Mx] & [\%/h]& [G] & [G] & [Mm] & [km/s] & [Mm] & 
[km/s] & [Mm] & [km/s] & [Mm]\\\hline\rule{0pt}{2ex}\hphantom{1}%
20\_opp000o   &   7.484 & 10.546 &  0.290 &  0.680 &  1.156 & 4275.1 & 2247.0 &  7.488 & $-$1.261 &   8.35 & $-$3.598 &   8.74 &  0.606 &  17.57\\\hphantom{1}%
40\_opp000o   &   6.975 & 11.029 &  0.368 &  0.676 &  3.655 & 5424.2 & 2213.2 &  5.376 & $-$1.076 &   7.49 & $-$2.662 &   8.16 &  0.676 &  14.30\\\hphantom{1}%
80\_opp000o   &   6.723 & 11.508 &  0.417 &  0.628 &  4.420 & 5833.2 & 2154.4 &  5.376 & $-$0.558 &   7.58 & $-$0.793 &   7.10 &  2.339 &  12.19\\
160\_opp000o  &   6.236 & 12.329 &  0.494 &  0.631 & 11.668 & 4346.9 & 2096.7 &  5.760 & $-$0.771 &   7.01 & $-$0.903 &   7.01 &  3.314 &  12.29\\
160\_opp000ho &   7.454 & 12.854 &  0.420 &  0.779 &  1.683 & 4518.3 & 2026.2 &  6.368 &         &         & $-$0.302 &   6.14 &  2.469 &  13.38\\
160\_opp000h  &   5.956 & 12.529 &  0.525 &  0.598 &  5.641 & 5360.4 & 2194.3 &  4.960 & $-$0.366 &   5.41 & $-$0.408 &   5.41 &  4.297 &  14.91\\
160\_opp050c  &   7.221 & 13.329 &  0.458 &  0.703 & 12.148 & 4838.8 & 2349.8 &  4.992 & $-$0.575 &   7.10 & $-$0.560 &   7.10 &  3.476 &  14.02\\
160\_opp100c  &   7.355 & 14.001 &  0.475 &  0.764 & 11.258 & 5239.6 & 2538.5 &  6.144 & $-$0.562 &   7.39 & $-$0.632 &   5.38 &  3.357 &  13.44\\
160\_opp150c  &   7.670 & 14.527 &  0.472 &  0.837 &  9.408 & 5652.8 & 2737.0 &  5.280 & $-$0.563 &   8.74 & $-$0.872 &   6.82 &  3.826 &  14.69\\
160\_opp200c  &   8.058 & 15.125 &  0.467 &  0.916 & 10.771 & 5924.1 & 2893.7 &  6.144 & $-$0.691 &   6.43 & $-$1.266 &   6.91 &  3.630 &  15.17\\
160\_opp300c  &   8.658 & 15.939 &  0.457 &  1.081 & 11.502 & 6607.6 & 3084.6 &  6.240 & $-$0.566 &   7.10 & $-$0.988 &   7.20 &  3.719 &  18.72\\
160\_opp000   &   6.789 & 12.975 &  0.477 &  0.648 &  8.468 & 4464.8 & 2166.0 &  5.856 & $-$0.578 &   6.24 & $-$0.740 &   6.24 &  3.045 &  14.59\\
160\_opp050   &   6.848 & 12.891 &  0.469 &  0.630 & 10.494 & 4478.0 & 2142.3 &  5.952 & $-$0.463 &   6.24 & $-$0.422 &   6.34 &  3.086 &  14.50\\
160\_opp100   &   6.776 & 12.861 &  0.473 &  0.608 & 10.745 & 4435.4 & 2261.7 &  5.760 & $-$0.589 &   7.30 & $-$0.895 &   6.62 &  3.709 &  13.25\\
160\_opp150   &   6.790 & 13.001 &  0.477 &  0.593 & 12.061 & 4380.4 & 2212.5 &  5.664 & $-$0.598 &   6.14 & $-$0.932 &   6.72 &  3.694 &  12.67\\
160\_opp200   &   6.638 & 12.921 &  0.486 &  0.577 & 10.392 & 4475.4 & 2199.3 &  5.952 & $-$0.724 &   6.53 & $-$1.168 &   6.53 &  3.625 &  12.77\\
160\_opp300   &   6.758 & 12.877 &  0.475 &  0.567 & 12.328 & 4360.5 & 2208.2 &  5.856 & $-$0.574 &   6.43 & $-$0.656 &   6.43 &  3.897 &  12.58\\
160\_opp100w  &   6.633 & 13.052 &  0.491 &  0.577 & 11.040 & 4239.6 & 2081.8 &  6.144 & $-$0.581 &   7.01 & $-$0.708 &   6.24 &  3.601 &  14.88\\
160\_opp150w  &   6.487 & 13.312 &  0.512 &  0.569 & 12.343 & 4241.1 & 2095.3 &  4.704 & $-$0.545 &   6.72 & $-$0.399 &   6.62 &  3.321 &  14.40\\
160\_opp200w  &   6.606 & 13.307 &  0.503 &  0.553 & 12.981 & 4282.6 & 2163.9 &  5.376 & $-$0.494 &   6.53 & $-$0.516 &   5.66 &  3.757 &  13.15\\
160\_opp000b  &   8.498 & 11.659 &  0.271 &  0.709 &  4.715 & 4543.3 & 1611.7 &  7.296 & $-$1.001 &   9.89 & $-$1.524 &  10.37 &  0.657 &  16.80\\\hline\rule{0pt}{2ex}%
alpha1.0      &  10.698 & 13.937 &  0.232 &  1.059 & $-$4.309 & 3641.2 & 1597.5 &  9.632&         &         &         &          &  1.050 &  11.26\\
alpha1.5      &  10.722 & 15.080 &  0.289 &  1.031 & $-$5.540 & 3877.7 & 2248.9 &  9.504&         &         &         &          &  3.985 &  13.34\\
alpha2.0      &  10.881 & 16.482 &  0.340 &  1.015 & $-$5.873 & 4180.3 & 2379.9 &  9.504&         &         &         &          &  4.130 &  14.40\\
alpha2.5      &  10.971 & 17.288 &  0.365 &  1.012 & $-$5.765 & 3359.9 & 2548.9 &  9.152&         &         &         &          &  4.380 &  14.43\\\hline\rule{0pt}{2ex}%
NOAA AR 11591 &   6.326 & 14.397 &  0.561 &  0.548 &        & 2829.2 & 1402.0 &  6.888 &&&&&\\\hline
\end{tabular}
\setlength{\tabcolsep}{\tmpSep}
\end{table*}}

\subsection{Initial condition: Potential with opposing flux}

The initial vertical component of the magnetic field at the bottom of the box is given by 
\begin{align}
  B_{\mathrm{z},0}(r)&=B_0\exp\left(-\frac{r^2B_0\pi}{F_\textrm{Gauss}}\right)-B_\textrm{opp}, 
\end{align}
where $B_0$ is the initial maximal magnetic field, $r$ is the distance from the spot axis, and $F_\textrm{Gauss}$ is the flux contributed by the \mbox{Gaussian} distribution. An alternative to a pure Gaussian is the subtraction of a uniform vertical magnetic field, $B_\textrm{opp}$. 
Performing a horizontal integration gives the total flux through the box,
\begin{align}
  F_\textrm{tot}&=\int B_{\mathrm{z},0} \left(\sqrt{x^2+y^2}\right)\mathrm{d}x\,\mathrm{d}y
  =F_\textrm{Gauss}-B_\textrm{opp}w^2,
\end{align}
where $w$ is the width of the box in both the horizontal directions. This value is the same for all heights and all times (but not over the $\tau=1$ iso-surface).
The initial field throughout the box is given by potential field extrapolation according to 
\begin{align}
B_\mathrm{z}&=\mathcal{F}^{-1}\left(\mathcal{F}(B_{z,0})                 \mathrm{e}^{-z\,|k|}\right)\,\text{ and }\,%
B_\mathrm{x}=\mathcal{F}^{-1}\left(\mathcal{F}(B_{z,0})\frac{-ik_\mathrm{x}}{|k|}\mathrm{e}^{-z\,|k|}\right),\label{eq:pot}
\end{align}
where $\mathcal{F}$ represents the horizontal Fourier transform. $B_\mathrm{y}$ can be obtained by replacing $k_\mathrm{x}$ by $k_\mathrm{y}$, or in our axisymmetric case, via $B_\mathrm{y}(x,y)=B_\mathrm{x}(y,x)$. This extrapolation is the same as the top boundary condition mentioned above, with the difference that there $z=0$ stands for the top (extrapolation into the ghost cells during the simulation run) instead of the bottom of the box (creating the initial field).

The potential extrapolation results in the initial $B_\mathrm{z}$ falling much more slowly with increasing radius outside the spot (see Fig.~\ref{fig:mag_ini}, top-left panel) than self-similar initial conditions as in \cite{2012ApJ...750...62R}.
For all initial conditions with a pure Gaussian in the smaller simulation box, $B_\mathrm{z}$ remained above 155\,G everywhere in the box. Given that in observations, $B_\mathrm{z}$ outside the spot vanishes, we tested the effect of subtracting a uniform vertical field ($B_\textrm{opp}$). This subtraction can be performed before or after extrapolation and the result is the same.

\myTabSims
\subsection{Simulation dataset}
The simulation runs described in this article are listed in Table~\ref{tab:sims}, where the following parameters were varied:
\begin{itemize}
\item Initial maximal magnetic field strength $B_0$: 20\,kG, 40\,kG, 80\,kG, or 160\,kG (most runs). $B_0$ is the first number in run names.\footnote{These unphysically strong initial fields at the bottom of the box get quickly reduced by an initial transient.}
\item Box width, $w$: 49.152\,Mm (most runs) or 98.304\,Mm. Runs with $w=98.304\,$Mm have names ending in `w'.
\item $F_\textrm{Gauss}=10^{22}$\,Mx (most runs) or $F_\textrm{tot}=10^{22}$\,Mx. Runs with $F_\textrm{Gauss}>10^{22}\,$Mx have names ending in `c'.
\item Opposing magnetic field $B_\textrm{opp}$: 0\,G, 50\,G, 100\,G, 150\,G, 200\,G, and 300\,G. This is the second number in the simulation names.
\item Most runs use a 96~(32)\,km horizontal (vertical) resolution, except the high-resolution runs with 32~(16)\,km, which have a name ending containing `h'. All low-resolution simulations use a grey radiative transfer.
\end{itemize}
The simulations with names ending in `o' were run with a previous version of the MURaM code and provided bolometric \mbox{intensities} instead of 500\,nm continuum intensities. They also contain further subtle differences that we are discussed below, where necessary. 

The box was 8.192\,Mm deep for all our simulations, but we provided comparison simulations `alpha*', which were only 6.114\,Mm deep. These are continuations of those started in \cite{2012ApJ...750...62R}, which were described in \cite{2020A&A...638A..28J} as `type II'. They have self-similar initial conditions and top \mbox{boundary} conditions, which force the field to be more horizontal than a potentially extrapolated field by a factor of $\alpha$.
\begingroup\linespread{0.98}\selectfont

While the simulation `160\_opp000o' continued running at low resolution, after 4.8121\,h, it was regridded to the higher 32~(16)\,km resolution; at the time 7.2163\,h, it switched over to a non-grey radiative transfer. In that form, which we labelled as `160\_opp000ho', it was described as `type I' in \cite{2020A&A...638A..28J}. 
Simulation `160\_opp000h' was run at the higher resolution from the start.

All high-resolution simulations presented here (`160\_opp000\{h,ho\}', `alpha*') use non-grey radiative transfer \citep{1982A&A...107....1N, Ludwig:1992,2004A&A...421..741V} with four opacity bins. The main benefit of this is that it allows forward radiative transfer, degradation to instrumental resolution, adding noise, and subsequent inversions, as done in \cite{2020A&A...638A..28J}.
The simulation `160\_opp000b' suppresses mass flows through the bottom in regions where $|B_\mathrm{z}|\gtrsim10\,$kG.
The azimuthal averages of the initial conditions at photospheric heights are shown in Fig.~\ref{fig:mag_ini}.
In addition, we used the Helioseismic and Magnetic Imager (HMI/SDO) \citep{2012SoPh..275..207S} observations of NOAA AR 11591 for comparison. For more details on this spot, in particular, its umbral and spot boundary, see \citet{2018A&A...620A.104S, outer}.

\subsection{Data processing}
During the simulation run, every 50 time steps, the intensity, $I_\textrm{c}$, as well as velocities, $v_\mathrm{xyz}$, and magnetic field vectors, $B_\mathrm{xyz}$, at the location of the $\tau=1$ layer were saved. This corresponds to one slice every $\approx$\,36\,s or $\approx$\,43\,s. The larger time steps are for the older simulations (with names ending in `o'), where the Alfv{\'e}n speed limiter was set to a lower value. For every simulation, we evaluated 101 such slices, thus covering $\approx$\,1\,h or $\approx$\,1.2\,h. The time ranges covered are listed as $t_\textrm{ref}$ in Table~\ref{tab:sims}. The different Alfv{\'e}n speed limiter values only change the length of the computation time steps and affect the low-density regions above the umbra, which are not relevant to this study.

The quiet Sun intensity, $I_\textrm{qs}$, as well as the umbral and spot boundaries, were calculated using the same methods as in \cite{2021A&A...656A..92S}, where the umbral boundary corresponds to $I_\textrm{c}=0.5 I_\textrm{qs}$ and the spot boundary to $I_\textrm{c}=0.9 I_\textrm{qs}$. For consistency, we calculated a single $I_\textrm{qs}$ for all new simulations (not ending in `o') by averaging over the quiet sun data of all simulations and all time steps and we did  the same, separately, for the old simulations (with names ending in `o'). These two sets of simulations were treated separately because the changes in MURaM between these runs resulted in incompat\-ible \mbox{intensity} values (500\,nm continuum vs bolometric).
Maps of the intensity and velocities indicate that the umbral boundary is sharp and well retrieved by this procedure, whereas an accurate determination of the spot boundary would require identifying individual convective cells and classifying them as belonging to the spot or quiet Sun. Using a contour on a degraded intensity map (as we do here) will only result in an approximate result, but it is good enough to retrieve the spot areas and radii.

In the next step, the vectors were transformed to obtain the radial field\footnote{Radial in sunspot simulations refers to a cylindrical coordinate system, with $z$ aligned with the spot axis and increasing with height.}, $B_\mathrm{r}$, total field, $|B|$, inclination to the surface normal $\gamma=\arctan\left(\sqrt{\smash[b]{B_\mathrm{x}^2+B_\mathrm{y}^2}}/B_\mathrm{z}\right)$ and corresponding results for the velocities. The values were then remapped from Cartesian to cylindrical coordinates, after which an azimuthal average was performed. 
In addition to the averages of the values themselves (e.g. mean vertical velocity, $v_\mathrm{z}$), the fraction of positive terms (upflow filling factor $n_{v_\mathrm{z}>0}$) and signed averages (mean upflow velocity, $v_{\mathrm{z},{+}}$, mean downflow velocity, $v_{\mathrm{z},{-}}$) were calculated.

In the final step, we performed a least-squares linear fit of all time-dependent parameters, including the azimuthal averages. Then we evaluate the value of this linear function at 4\,h (`160\_opp000ho': 7.5\,h, `alpha*': $6.\bar{3}\,$h). For simplicity, hereafter in this article, these temporal fits are implied unless mentioned otherwise. For instance, when we refer to azimuthal aver\-ages, we mean these values of the linear functions at 4\,h fit to the azimuthal averages.
When we refer to maps in this work, we mean those of the last time in the reference range (last column of Table~\ref{tab:sims}).

\myTabRes\begin{figure}[t]
\includegraphics[width=1\columnwidth]{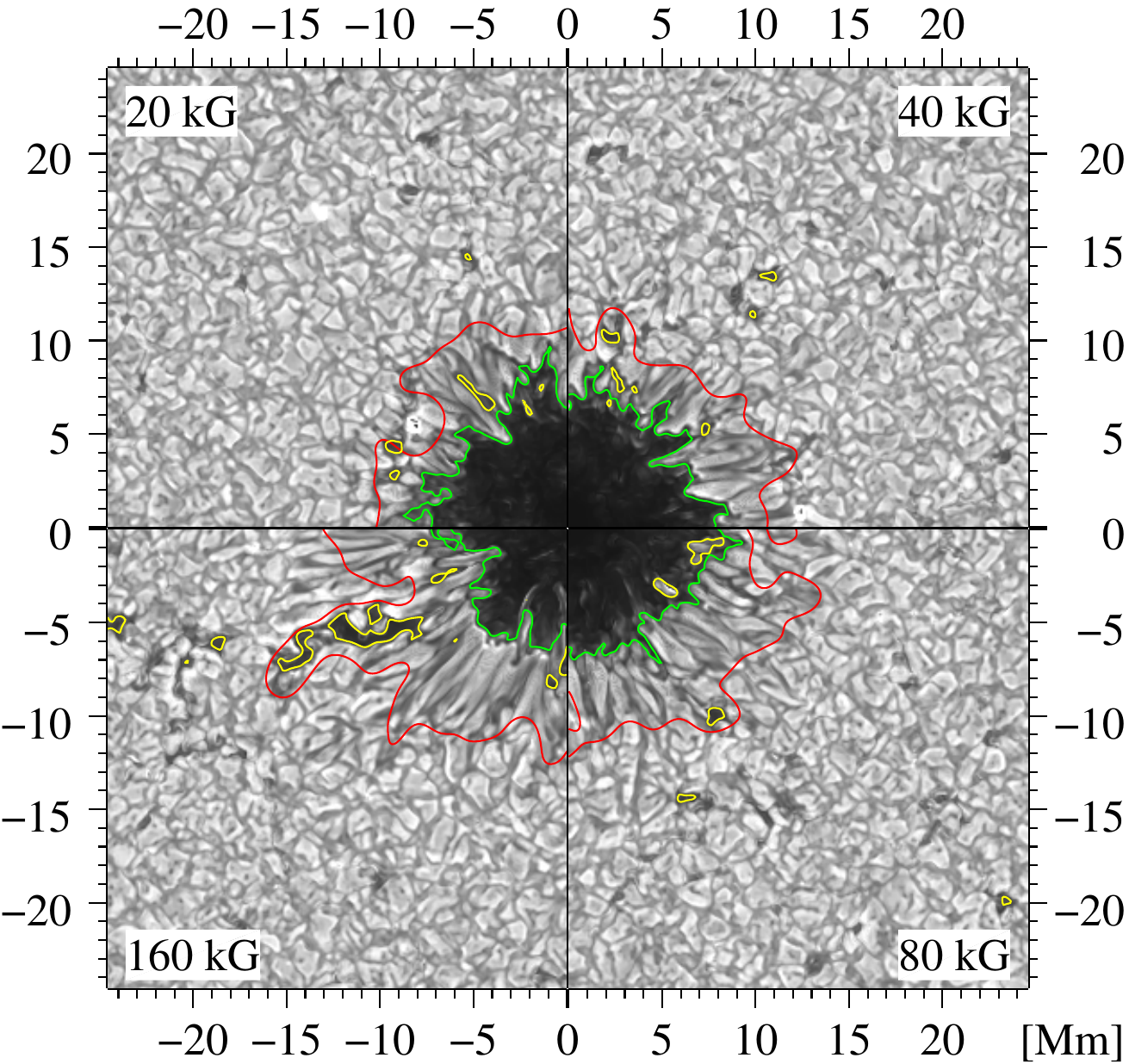}
\caption{Quadrants of bolometric intensity maps for four selected simulation runs with different initial magnetic field strengths. From top-left clock-wise: $B_0=20$\,kG, 40\,kG, 80\,kG, and 160\,kG. The magnetic flux is the same for all four runs: $F=10^{22}$\,Mx.}
\label{fig:I_out_map}
\end{figure}
\begin{figure*}[t]
\includegraphics[width=\linewidth]{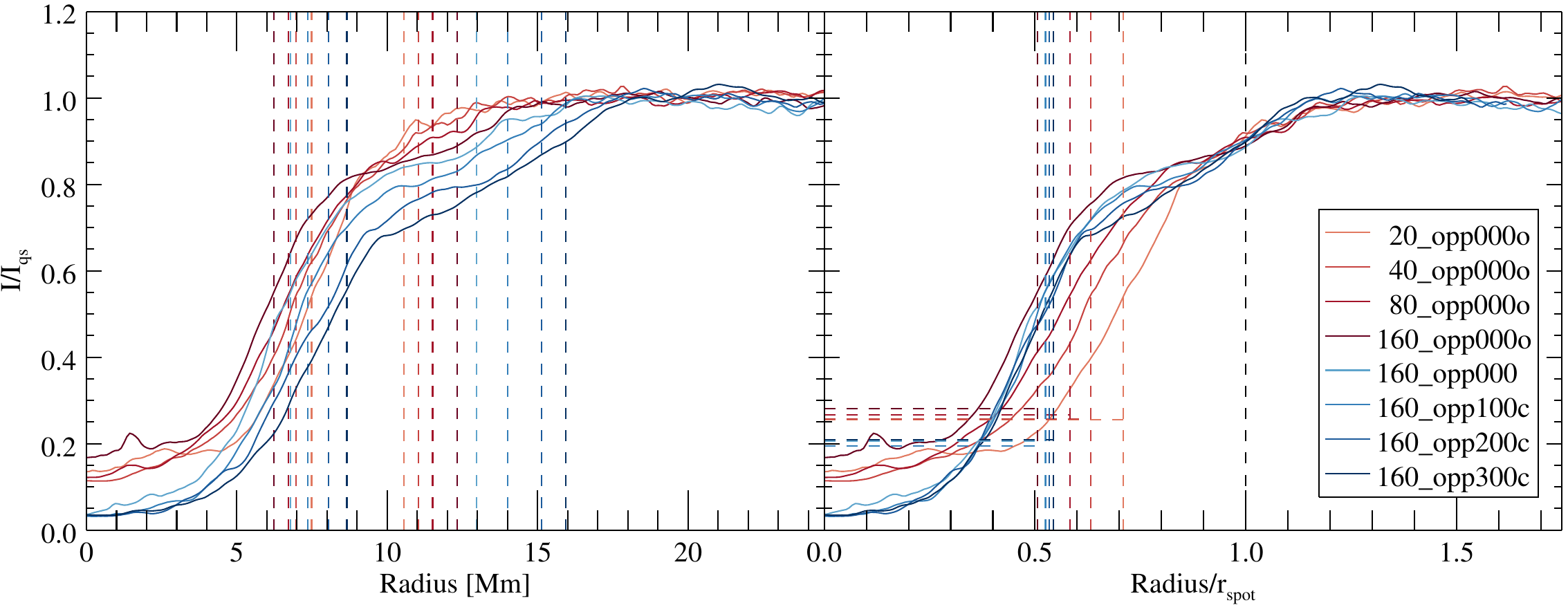}
\caption{Azimuthally averaged intensity, $I/I_\textrm{qs}$, (solid lines) as a function of radius in Mm and fraction of the spot radius. The colours indicate the simulations listed in the inset legend. 
For the older runs (with names ending in `o' in red), the bolometric intensity is shown, and for the other runs (blue), the continuum intensity is shown.
Vertical dashed lines show the position of the umbral (left panel with $R<10$\,Mm, right panel $R<1$) and spot boundaries (left panel $R>10$\,Mm, right panel $R=1$ black). The horizontal dashed lines in the right panel show the average umbral intensity.
Values corresponding to the location of the vertical dashed lines are given in Table~\ref{tab:res}, columns $r_\textrm{u}$, $r_\textrm{s}$, $r_\textrm{u/s}$.
}
\label{fig:I_out_r}
\end{figure*}
\section{Results}
\label{sec:results}
\subsection{Intensity}
Figure~\ref{fig:I_out_map} shows the bolometric intensity maps of sunspot simulations with four values of the initial maximal magnetic field strength $B_0$: 20\,kG, 40\,kG, 80\,kG, and 160\,kG, respectively.  All runs contain the same flux (10$^{22}$\,Mx), with no opposing field $B_\textrm{opp}$, and situated in the small (49\,Mm wide) box. The green line shows the umbral boundary as the longest contour of $I=0.5I_\textrm{qs}$, in yellow other contours at the same intensity level, and in red the spot boundary as the longest contour at $I_\textrm{c}=0.9I_\textrm{qs}$. The other contours at this level are not shown.

As a global property, the size of the penumbrae increased with increasing initial magnetic field strengths, $B_0$, up to 160\,kG.
Simulation runs with $B_0=20\,$kG and $40$\,kG show elongated convective cells resembling penumbral filaments. The $B_0=160\,$kG run shows the slenderest and longest penumbral filaments, and $B_0=80\,$kG shows an intermediate state.

\endgroup
\hereOrThere{\myFigBzAzi}{\begin{figure}[!htb]
\includegraphics[width=1\columnwidth]{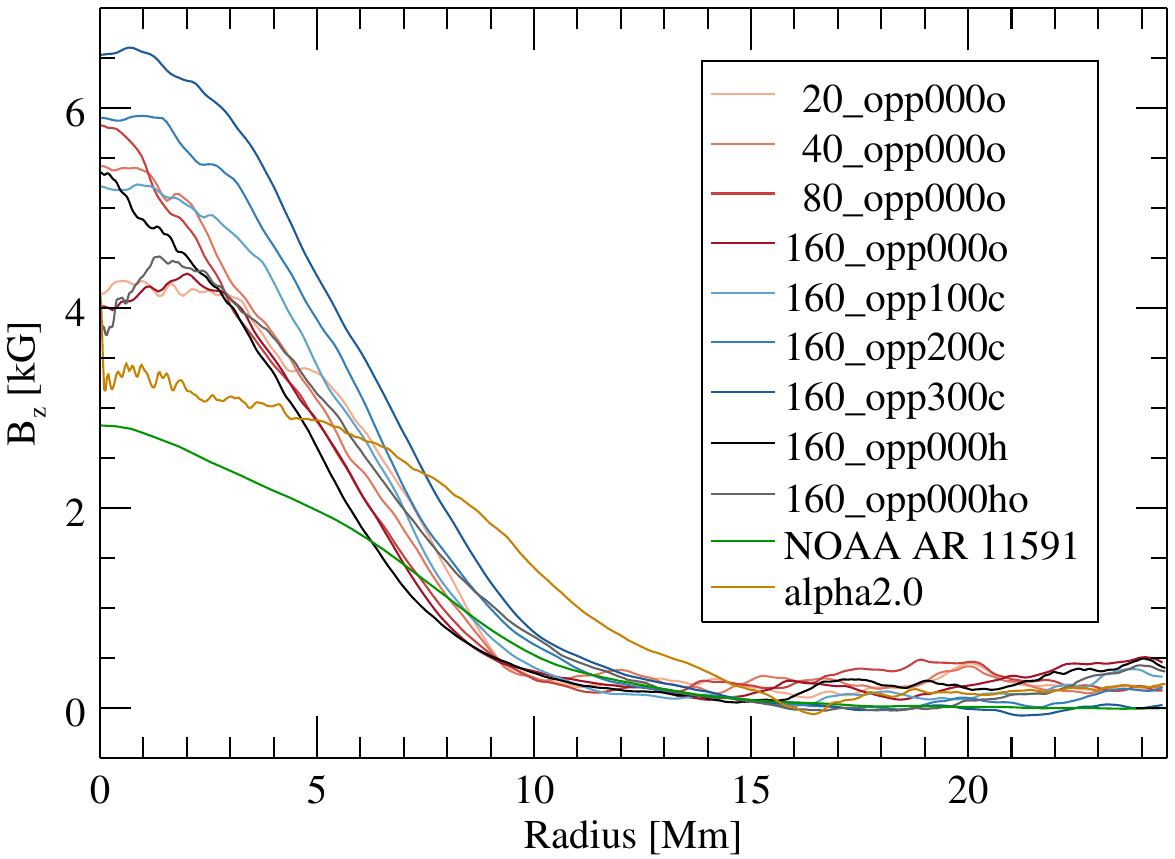}
\caption{Vertical magnetic field component, $B_\mathrm{z}$, of selected simulations. The $B_\mathrm{z}$ value from an observation (green) is also displayed for comparison.}
\label{fig:Bz_azi}
\end{figure}}
Figure~\ref{fig:I_out_r} and the bottom-right panel of Fig.~\ref{fig:mag_azi} show the \mbox{azimuthal} averages of the intensity for selected simulations. The vertical dashed lines show the average spot and umbral radii, $r$, calculated from the contour areas using $r=\sqrt{A/\pi}$. For all simulations, the umbral and spot sizes, along with their penumbral fractions are given in Table~\ref{tab:res} (columns $r_\textrm{u}$, $r_\textrm{s}$, and $r_\textrm{pu}/r_\textrm{s}=1-r_\textrm{u}/r_\textrm{s}$). 
Increasing $F_\textrm{Gauss}$ beyond $10^{22}\,$Mx increases the sizes of the umbrae and spots.
Based on sunspot statistics, \cite{Kiess+2014} showed that the umbral radii increase linearly with the spot flux. In our simulations, the umbral radii increased with spot flux, consistent with their linear fit.
However, for $B_0 < 160\,$kG, the penumbral fractions are smaller than for those with $B_0=160\,$kG, which, at 47\%–52\%, are already below the typical value observed in sunspots, where the penumbra covers about 60\% of the spot size 
\citep[e.g.][and references therein]{Keppens:1996,cwp:2001, Mathew:2003, Borrero:2004, Bellot:2004, Sanchez:2005, Beck:2008, Borrero:2011, 2024SoPh..299...19C}.
In summary, our simulations provide sunspots with a realistic umbral size but diminished penumbral length; that is, our simulations provide smaller sunspots than the observed stable ones. However, the simulated spot sizes resem\-ble those of developing sunspots during penumbra formation \citep[see e.g.][]{Jurcak:2014}.

The difference in brightness in the umbrae between older runs (names ending in `o', red lines in Figs.~\ref{fig:I_out_r}) and the newer ones (blue lines) are caused by the former giving bolometric \mbox{intensities}, $I=I_\textrm{bol}$; whereas the latter are giving the contin\-uum intensity at 500\,nm, $I=I_{c500}$. The differences between these intensities are consistent with the expectations of black\-body \mbox{radiation}.

\subsection{Surface magnetic fields}
\myFigBzAzi
\begin{figure*}[!htb]
\includegraphics[width=\linewidth]{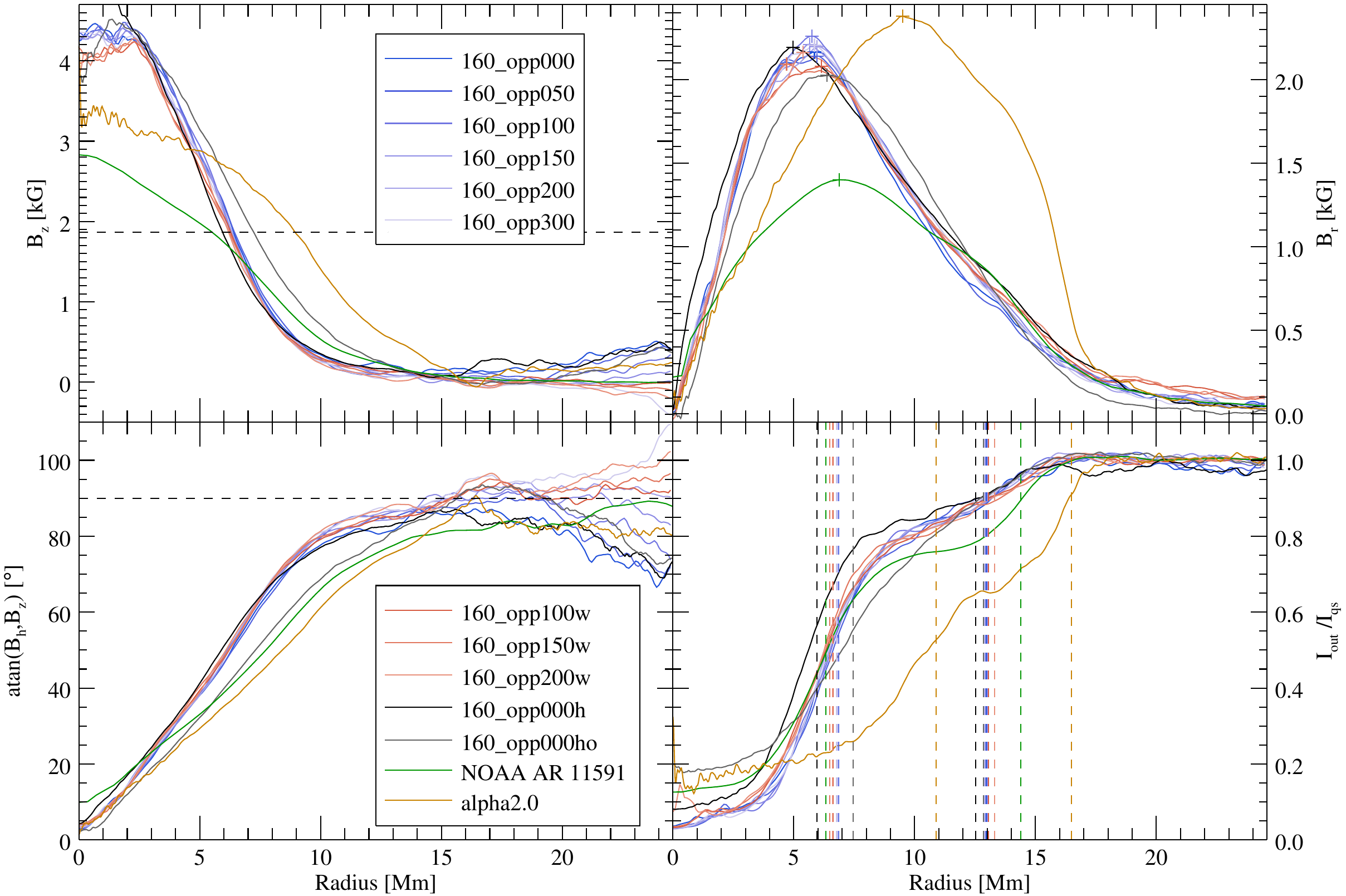}
\caption{Azimuthal averages of the vertical magnetic field component, $B_\mathrm{z}$ (top left), the radial magnetic field component, $B_\mathrm{r}$ (top right), the inclination (bottom left), and the intensity, $I_\textrm{out}$ (bottom right), of our more realistic simulations (blue, red, black and grey), and of an observation (green), and an old reference simulation (gold). Horizontal dashed lines refer to $B_\mathrm{z}=1867\,$G, the observed critical value for penumbral-type convection to operate (top left), and to an inclination of $90\degr$, meaning a horizontal field (bottom-left).}
\label{fig:mag_azi}
\end{figure*}

In Fig.~\ref{fig:Bz_azi}, we plot the radial dependence of the azimuthally aver\-aged $B_\mathrm{z}$ for the runs given in the figure’s legend.
Simulations with $F_\textrm{Gauss}>10^{22}\,$Mx (names ending in `c') have stronger vertical magnetic fields in the spot axis $B_\mathrm{z}(r=0)$ at the average $\tau=1$ height in the initial condition (see top-left panel of Fig.~\ref{fig:mag_ini}, three light blue lines) than those with $F_\textrm{Gauss}=10^{22}\,$Mx (dark blue line). This results in unrealistically strong fields in the later stages of the simulation (see Fig.~\ref{fig:Bz_azi}, blue lines, and the `$\max B_\mathrm{z}$' column in Table~\ref{tab:res}).
By unrealistic, we mean that they are stronger than observed on the sun in spots of comparable fluxes \citep{2012A&A...541A..60R,Kiess+2014}.

The initial phase of the simulations had both concentrating and dispersing effects in the magnetic fields at the spot centre. 
For the $B_0=160\,$kG simulations starting with the lower resolution (96\,/\,32\,km), the concentrating effects are slightly stronger than the dispersing effects, \mbox{resulting} in a moderate \mbox{increase} in $B_\mathrm{z}$. For the simulation starting with a higher resolution (`160\_opp000h'), the concentrating effects were stronger, resulting in an excessively high $B_\mathrm{z}$.
For $B_0=40\,$kG and 80\,kG, the dispersing effects are weaker than those for the $B_0=160\,$kG simulations, resulting in an overly strong $\max B_\mathrm{z}$.
This is consistent with a simulation with $B_0=160\,$kG, but with a closed bottom boundary at the spot foot-point (`160\_opp000b'), which also lacks the very dynamic initial phase with its dispersion effects and has a higher $\max B_\mathrm{z}$. The simulation with a closed bottom boundary also lacked a properly formed penumbra and the few elongated granules had inflows along the full length and downflows outside the umbral boundary.

Low-resolution simulations with $F_\textrm{Gauss}=10^{22}\,$Mx and $B_0=160\,$kG have $\max B_\mathrm{z}=4240{-}4478\,$G, while $B_\mathrm{z}$ drops with \mbox{increasing} radius to values of $200\,$G similar for all such simulations. Outside the spot, the values for $B_\mathrm{z}$ fluctuate, whereas simulations in larger boxes or higher $B_\textrm{opp}$ fluctuate around lower values, as shown in the top left panel of Fig.~\ref{fig:mag_azi}.

Simulations with $F_\textrm{Gauss}>10^{22}\,$Mx have increased $B_\mathrm{r}$ compared to those with $F_\textrm{Gauss}=10^{22}\,$Mx. Beyond that, within the spot, all simulations with $B_0=160\,$kG have a similar dependence of $B_\mathrm{r}$ on the fractional spot radius, as depicted in the \mbox{upper} right panel of Fig.~\ref{fig:mag_azi}. The $B_0=20\,$kG simulation has the maximum $B_\mathrm{r}$ at a larger radius, consistent with the higher umbral fraction. All $B_0<160\,$kG simulations have a faster drop of $B_\mathrm{r}$ outside the spot than the $B_0=160\,$kG simulations. Of the $B_0=160\,$kG simulations, $B_\mathrm{r}$ decreases more slowly outside the spot for those in larger boxes, indicating that the drop-off of $B_\mathrm{r}$ in smaller boxes may be caused by the horizontal periodic boundary conditions.

The increase in field inclination, $\gamma$, with fractional spot \mbox{radius} is similar for all $B_0=160\,$kG simulations up to $\gamma\approx80^\circ$ (Fig.~\ref{fig:mag_azi}, bottom left panel), whereas $\gamma$ rises more slowly with the fractional spot radius for the $B_0<160\,$kG simulations, which is consistent with their larger umbral fraction. The other differences in the initial conditions are shown in the bottom right panel of Fig.~\ref{fig:mag_ini} within the spot were removed during the dynamic initial phase. Outside the spot, the average inclination fluctuated around $75\degr{-}105\degr$, depending on the average $B_\mathrm{z}$ there.

\subsection{Surface flows}
\label{sec:flows}
\begin{figure}
\includegraphics[width=1\columnwidth]{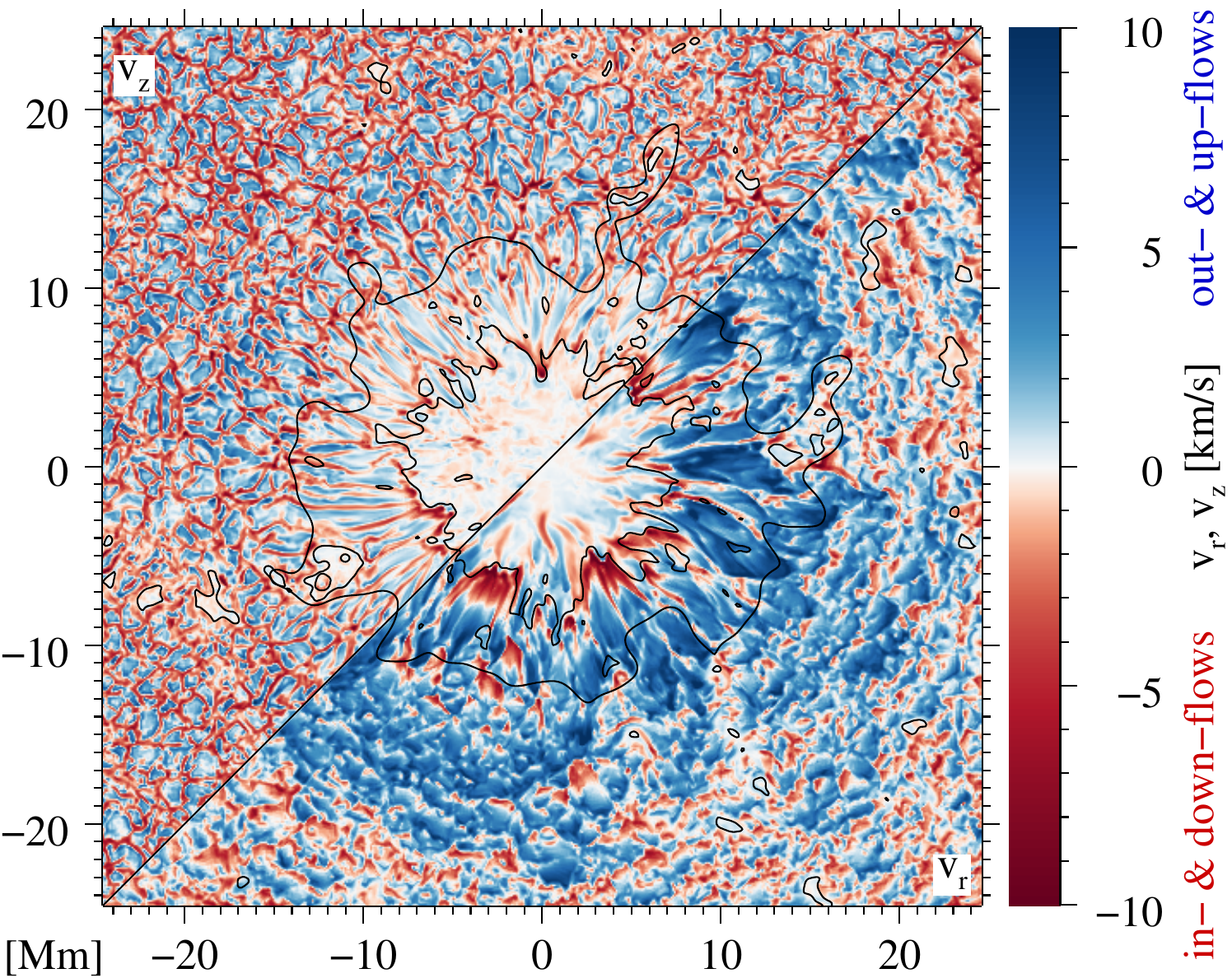}
\caption{Surface velocity map of sunspot run `160\_opp000'. Top-left: Vertical flows, $v_\mathrm{z}$, downflows in red, upflows in blue. Bottom-right: Radial velocities, $v_\mathrm{r}$, inflows in red (towards the spot centre), outflows in blue. The black contours are the same as shown in Fig.~\ref{fig:I_out_map}.} 
\label{fig:vr_map}
\end{figure}
\begin{figure}
\includegraphics[width=1\columnwidth]{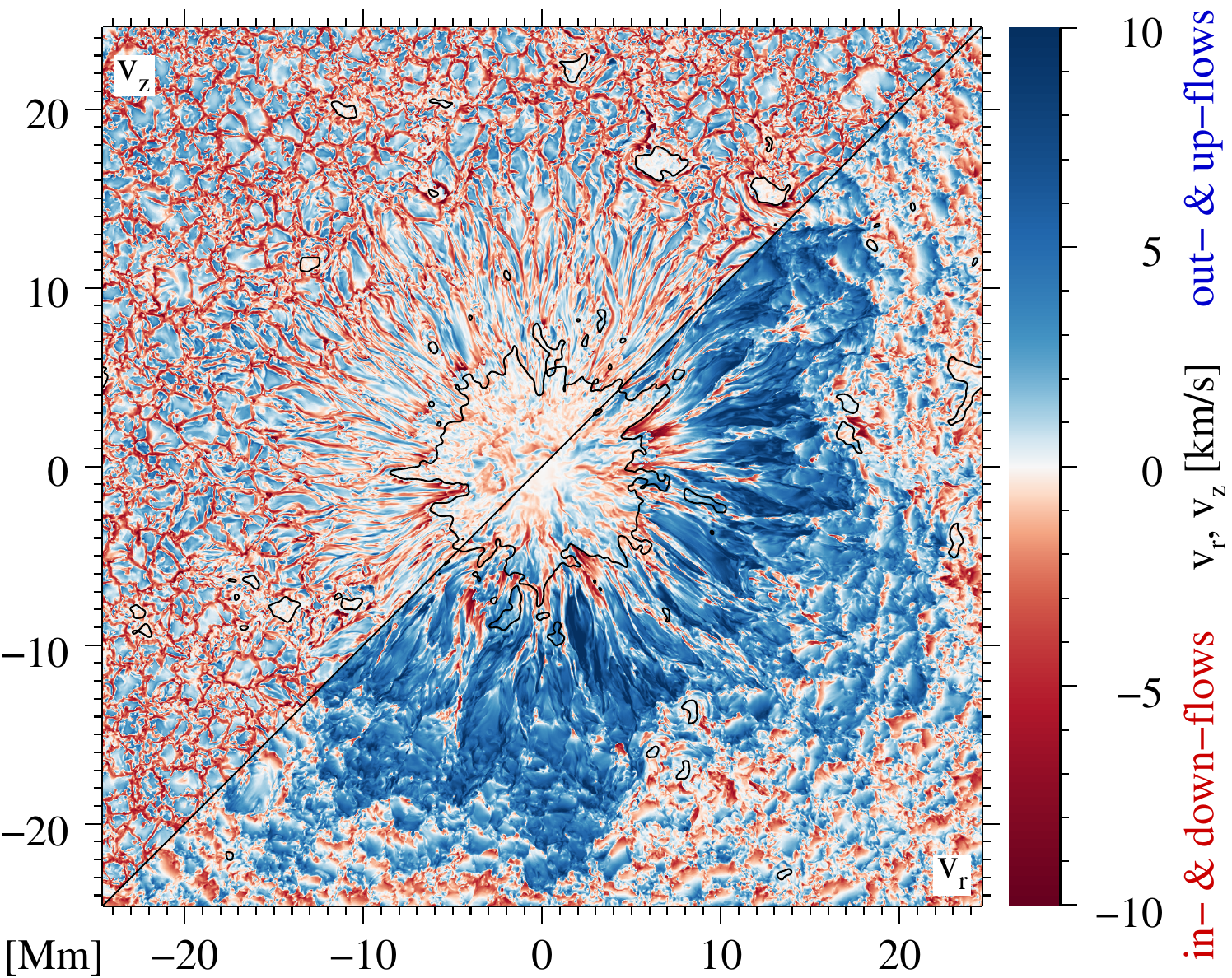}
\caption{Same as Fig.~\ref{fig:vr_map}, but for sunspot run `160\_opp000h' (high res.), with no spot boundary contour.}
\label{fig:vr_map_hr}
\end{figure}
\begin{figure}
\includegraphics[width=\columnwidth]{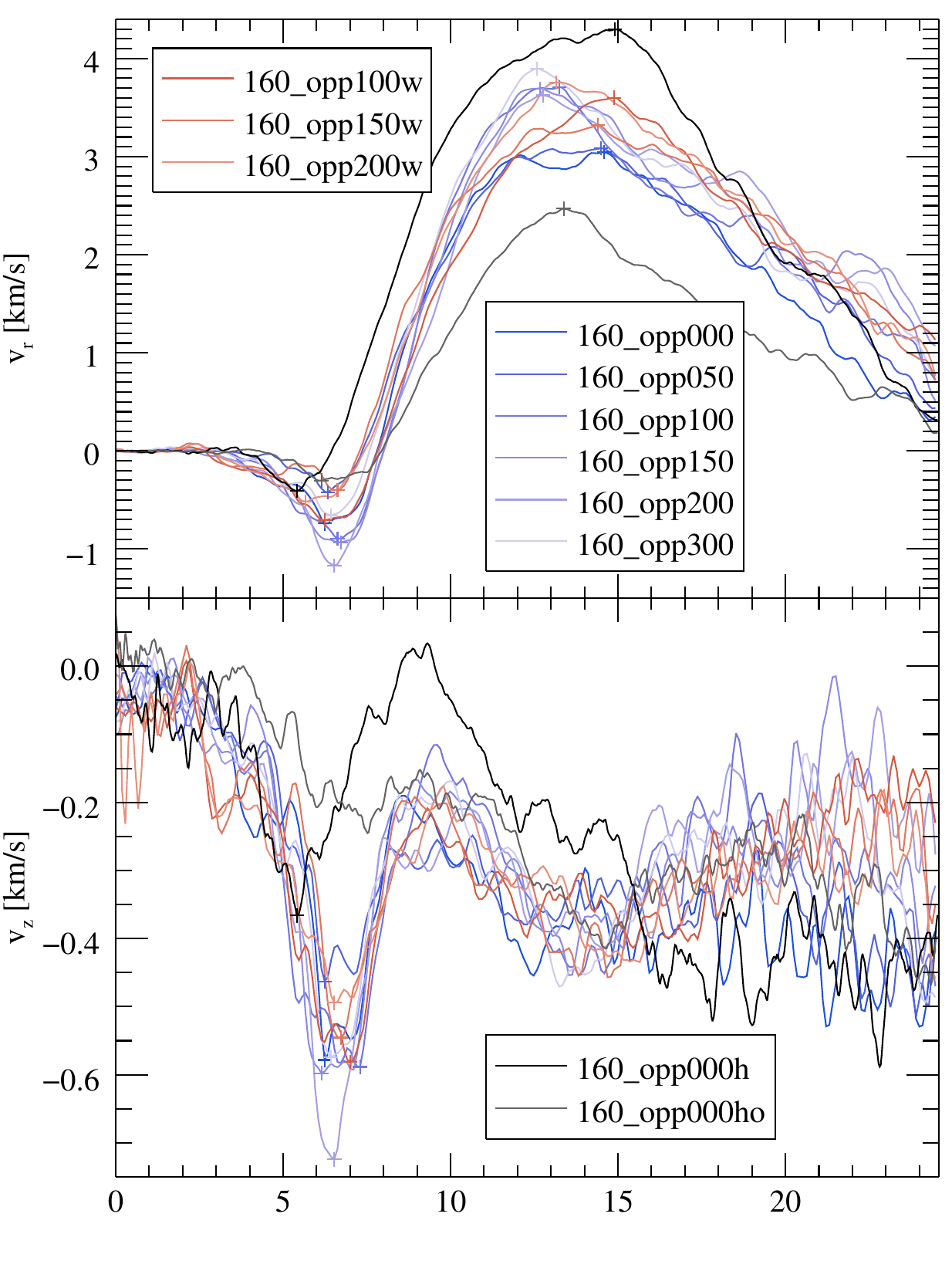}
\caption{Azimuthally averaged radial, $v_\mathrm{r}$ (top panel), and vertical, $v_\mathrm{z}$ (bottom panel) velocities of our more realistic simulations}
\label{fig:v_azi}
\end{figure}
The simulations with $B_0=20\,$kG and 40\,kG have inflows \mbox{directed} towards the umbra along the full length of all elongated convective cells, with downflows at umbral boundaries. Since this is clearly inconsistent with what is observed on the sun, we did not emark on a more detailed discussion of the surface flows in these  runs. A comparison of the other runs with observations is given in Sect.~\ref{sec:conclusions}.

Simulations with $B_0=80\,$kG and 160\,kG and low resolution (96/32\,km) have strong in- and downflows in most umbral ends of penumbral filaments (filament heads), with out- and upflows in the middle and outer parts of the filaments. Downflows were also observed between the filaments and at the outer ends. 
An example map of the radial flows (in- and outflows) and vertical flows (up and down) is presented in Fig.~\ref{fig:vr_map}.

In the umbra, the dominant contribution to the vertical flows, $v_\mathrm{z}$, is an overall oscillation, which at the time shown here is near a zero-crossing and, hence, it is not visible in the map. Our \mbox{azimuthally} averaged and temporally fitted values of $v_\mathrm{z}$ have these oscillations averaged out over time. This was verified in some simulations by subtracting the $\tau=1$ surface oscillations from the vertical velocities.
Outside the umbra, the averages of the upflow filling factor and separate averages over up- and downflow speeds change quantitatively, but the radial distance of the minima and maxima is not affected. Inside the umbra, these values are not usable because vertical oscillations are the dominant contributors to the velocity.

The following discussion focuses on the common flow \mbox{behaviour} of the $B_0=160\,$kG simulations. We find that the \mbox{simulation} run with $B_0=80\,$kG is similar in many aspects, but the shorter penumbral filaments result in slower outflows.

In terms of azimuthal averages, the outflow speed $v_{r,{+}}$ \mbox{increases} with increasing radius until the spot boundary and then slowly decreases. 
Inflow speeds of $-v_{r,{-}}$ increased until the \mbox{umbral} boundary, then converged to the lower quiet sun value.
The outflow filling factor, $n_{v_\mathrm{r}>0}$, has values of 39--48\% at a \mbox{local} minimum near the umbral boundary, then rapidly increases and tapers off to a maximum around the spot boundary of 88--93\% for the $B_0=160\,$kG simulations.
This results in the radial veloc\-ity, $v_\mathrm{r}$, having a minimum near the umbral boundary and a maximum near the spot boundary.
The minima, maxima, and their \mbox{locations} in our simulation runs are listed in Table~\ref{tab:res}, with $v_\mathrm{r}$ and $v_\mathrm{z}$  shown in Fig.~\ref{fig:v_azi}, where the extrema are marked with \mbox{$+$ symbols}.

Upflow speeds, $v_{z,{+}}$, increased steadily with radius until they reached the quiet-sun average at around $1.2\,r_\textrm{s}$.
Downflow speeds, $-v_{z,{-}}$, have a local speed maximum near the umbral boundary, followed by a decrease in speed over the inner penumbra to increase to the quiet Sun average.
Outside the umbral boundary, the upflow filling factor, $n_{v_\mathrm{z}>0}$, increased over the same range as the downflows slowed with increasing radius. Outside this narrow range, any trend in the radial dependence of the filling factor is sufficiently weak to be hidden by any fluctuations that might appear random.
The lower panel of (`160\_opp000h') shows that inside the central umbra, vertical velocities, $v_\mathrm{z}$, are, on average, near zero, then drop to a local minimum near the umbral boundary. This is followed by a quick increase over the inner part of the penumbra and a decrease to the quiet sun average over the width of the rest of the penumbra, accompanied by an increase in fluctuations.

Having a negative average quiet-sun vertical velocity does not stand in contradiction to the convective blue shift, since the convective blue shift results from weighting the average of the light, meaning that the bright granular upflows outshine the darker inter-granular downflows. Our average is negative \mbox{because} the $\tau=1$ surface in the inter-granular lanes is lower; hence, the downflows are faster than the upflows.

The maximal total velocity, $v$, was found to be outside, but near the spot boundary, beyond which it dropped to the quiet-sun average.
Most simulation runs have a minor local maximum near the umbral boundary.

In summary, as shown in Fig.~\ref{fig:vr_map}, a pattern that appears \mbox{characteristic} of the potential field simulations with $B_0\ge80\,$kG presented here is manifested by the bi-directional flows \mbox{exhibited} by the penumbral structure: radial outflows (blue) \mbox{resembling} the observed Evershed flows are present in the mid and outer penumbra, yet the inner penumbra is char\-acterised by strong inflows (red) combined with downflows. \mbox{Bi-directional} flows were first reported in high-resolution obser\-vations by \citet{2011IAUS..273..134S} as counter-Evershed flows. \mbox{Recently}, \citet{GarciaRivas+2024} showed that some of these counter-Evershed flows correspond to bi-directional flows that act as a precursor to the formation of a stable penumbra. The simulation with a closed bottom boundary inside the magnetic trunk (`160\_opp000b') has elongated granules instead of penumbral filaments and, hence, no outflows.

Figure~\ref{fig:vr_map_hr} shows that the high-resolution (32/16\,km) simulations have, however, both filaments with bi-directional flows and those with normal Evershed flows, meaning upflows in the filament heads, outflows along the whole lengths of the filaments, and downflows at the outer ends and between the filaments.
The simulation that started with high-resolution (`160\_opp000h') has faster outflows (Fig.~\ref{fig:v_azi}, upper panel), consistent with the resolution dependence of the Evershed flow speed reported by \citet{2012ApJ...750...62R}. The simulation that started at low resolution and was later regridded to high resolution (`160\_opp000ho') has slower outflows, because the flow speeds decrease over time and this simulation is evaluated at a later time.

\subsection{Flux emergence}
Our low-resolution (96/32\,km) simulations with $B_0=160\,$kG and open bottom boundary (names starting with `160' and an ending  that does not contain `h' or `b') keep increasing their flux during our observation time (Table~\ref{tab:res}, column $\mathrm{d}_tF/F$, 8.5--13.0\%/h). This corresponds to flux emergence. In the only simulation run for substantially longer than 4\,h (`160\_opp000'), the stalling of flux emergence at 8--10\,h coincides with a narrowing of the penumbra and the disappearance of most outflows.

In simulation `160\_opp000ho', for which the data were evaluated 3.5h later than in most other simulations, the flux emergence appears to be tailing off. 
In the simulation with a closed bottom boundary inside the magnetic trunk (`160\_opp000b'), only slight flux emergence is observed.

\section{Summary and conclusion}
\label{sec:conclusions}

As the aim of this study is to explore more realistic sunspot simulations by extending the ideas introduced by \cite{Nordlund2015} and examined in \cite{2020A&A...638A..28J}, we carried out a parameter study in MURaM sunspot simulations under the potential-field initial condition. Here, we varied the following parameters:
\vspace{-6pt}
\begin{itemize}
    \item Initial magnetic field strength, $B_0$: 20, 40, 80, and 160\,kG;
    \item Magnetic flux content $F_\textrm{Gauss}$ and opposing flux, $B_\textrm{opp}$;
    \item Box width;  
    \item Resolution (96\,/\,32\,km versus 32\,/\,16\,km).
\end{itemize}
\vspace{-6pt}
\noindent Our main results are:
\vspace{-6pt}
\begin{enumerate}
    \item The penumbral extent increases systematically with $B_0$, from no penumbra at 20\,kG to long, slender filaments at 160\,kG.
    \item Runs with $F_\textrm{Gauss} > 10^{22}$\,Mx develop unrealistically large umbral vertical fields, $B_\mathrm{z}$.
    \item \label{item:B_dist}The most realistic sunspots come from $B_0 = 160$\,kG, $F_\textrm{Gauss} = 10^{22}$\,Mx, which produce (1) maximum $B_\mathrm{z} = 4.2{-}4.5$\,kG (still larger than found in observations) and (2) a realistic profile of $B_\mathrm{z}$ and $B_\mathrm{r}$ with the radius.
    \item \label{item:close}Closing the bottom boundary inside the spot trunk does not create a realistic sunspot.
    \item Varying $B_\textrm{opp}$ or the box width changes the background $B_{z}$ outside the spot but does not significantly affect the internal spot structure or dynamics.
    \item For $B_{0} \leq 40$\,kG, no Evershed flows develop; there are only in- and downflows, with only slight flux emergence.
    \item \label{item:summary_LR}For $B_{0} = 80{-}160$\,kG and low resolution (96/32\,km), bi-directional flows, resembling those found in observations of the early stages of penumbra formation, are found during ongoing flux emergence.
    \item \label{item:summary_HR}High-resolution (32/16\,km) runs produce both filaments with bi-directional flows and filaments with Evershed flows.
\end{enumerate}
\vspace{-8pt}
\noindent From this parameter study, we conclude that simulations initialised with potential-field conditions, sunspot fluxes on the order of 10$^{22}$\,Mx, and strong (160 kG) magnetic fields and open bottom boundary reproduce key aspects of the \mbox{observed} mor\-phology and dynamics of sunspots during the early phases of penumbra formation \citep[see e.g.][]{Schlichenmaier:2010a, Jurcak:2014}. Specifically, our simulations have early penumbral-like regions characterised by slender granulation of reduced intensity; rather than exhibiting the classical \mbox{radial} velocities of the Evershed flow \citep{2000A&A...358.1122S,2013A&A...551A.105L}, they show bi-directional motions that are likely driven by flux emergence at the outskirts of the spot \citep{GarciaRivas+2024}. These results highlight the value of these simulations in investigating the initial stages of sunspot formation.

Comparisons with failed simulations show that -- at least for low-resolution simulations with potential field top boundary conditions -- we only get a 
realistic-looking 
sunspot with outwards flow in a sufficiently wide penumbra, while flux keeps emerging in the outer parts of the spot.\footnote{The causes behind the flux emergence in our simulations relate to the unphysical entropy enhancement during the initial transient and are beyond the scope of this article.}
We speculate that any low-density plasma in $\cap$-shaped magnetic field lines below the surface might be sufficient to drive flux emergence and create penumbrae as in our simulations.

However, observations suggest that an overlying canopy is required to form a stable penumbra with Evershed flows \citep{Lindner+2023, Chifu+2025}. Older simulations have achieved this by forcing the field to be more inclined at the top boundary, so that the inclined field drives the Evershed flows. 
The presence of Evershed flows in some of the filaments of our high-resolution simulations, coupled with their absence in low-resolution simulations, indicates that insufficient resolution is an inhibiting factor for Evershed flows, in line with preliminary \mbox{reports} from \citet{2025ghh..confE..13R}.
We speculate that embedding the spot in a much larger simulation box (across all dimensions above and below the photosphere), while including a small-scale dynamo run that allows a canopy to form naturally, might also force the field to be sufficiently inclined. As a result,  Evershed flows might indeed form even in low-resolution simulations if the spot is placed in the right location.

\bigskip
\begin{acknowledgements}
We thank the anonymous reviewer for their valuable and insightful comments, which helped improve this manuscript.
This work was supported by the Czech-German common grant, funded by the Czech Science Foundation under project 23-07633K and by the Deutsche Forschungsgemeinschaft under project BE 5771/3-1 (eBer-23-13412), and the institutional support ASU:67985815 of the Czech Academy of Sciences.
Most of the \mbox{simulations} leading to the results obtained were run on Piz Daint at the Swiss National Super\-computing Centre, Switzerland, financed through the ACCESS programme of the SOLARNET project, which has received funding from the European Union Horizon 2020 research and innovation programme under grant agreement no 824135.
For the older simulations, we would like to thank Matthias Rempel for providing us with their data and acknowledge the high-performance computing support from Cheyenne (doi:10.5065/D6RX99HX) provided by the NFS NCAR's Computational and Information Systems Laboratory, sponsored by the National Science Foundation. 
This research has made use of NASA's Astro\-physics Data System Bibliographic Services.
\end{acknowledgements}

\bibliographystyle{aa_url}  
\bibliography{biblio}
\sethyphenpenalty{50}
\onecolumn
\begin{appendix}
\section{Initial magnetic field configurations}
\xdef\thelastlineno{\the\numexpr\value{linenumber}+1\relax} 
\begin{figure*}[h]
    \centering
    \includegraphics[width=\linewidth]{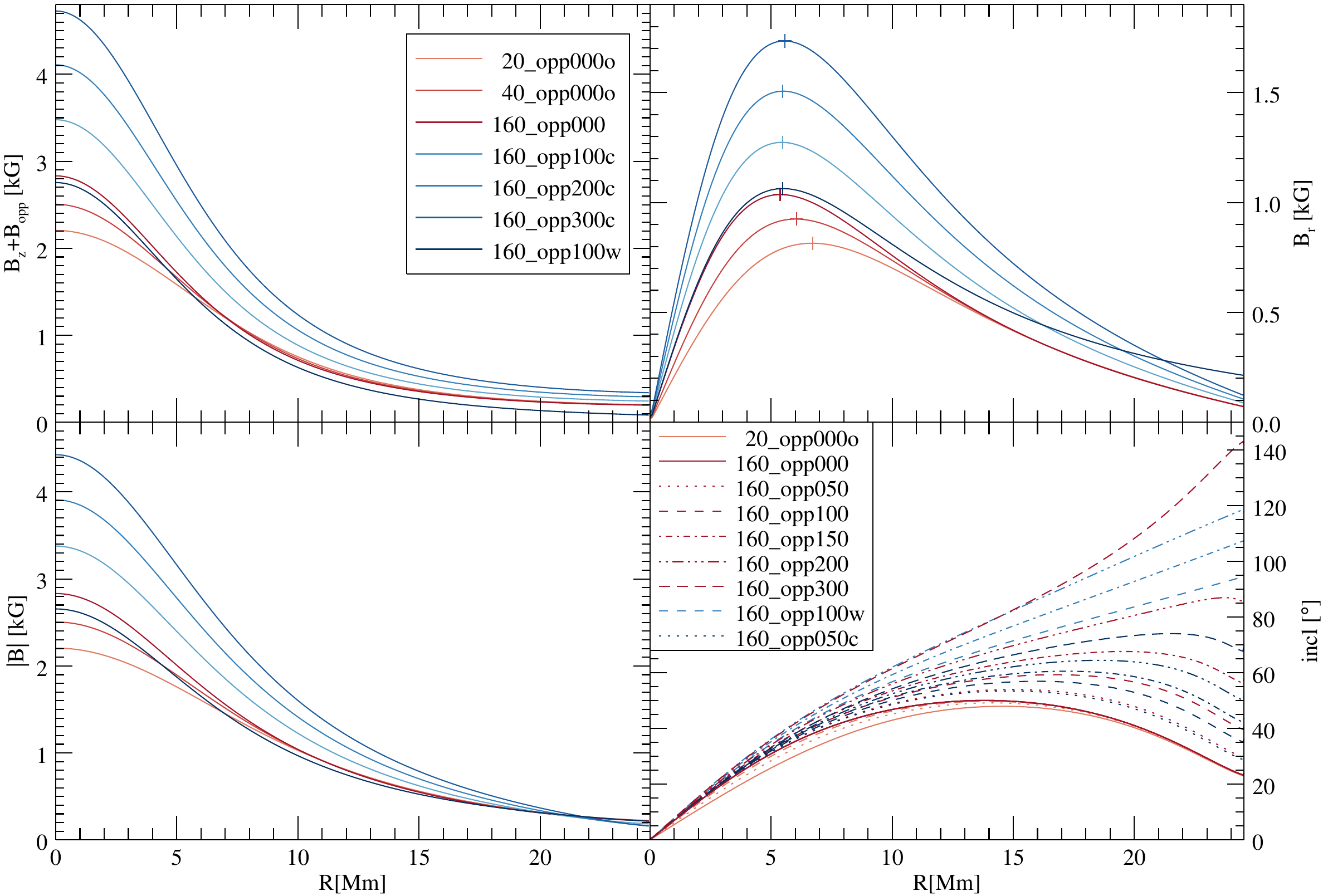}
    \caption{Initial magnetic field.  Vertical field $B_\mathrm{z}+B_\textrm{opp}$ (top left), radial field $B_\mathrm{r}$ (top right), total field $|B|$ (bottom left), and inclination to the surface normal $\gamma$. The legend in the top left panel shows the simulations plotted in the first three panels. The legend in the bottom right panel gives a selection of the simulations shown there, the rest is given in Table~\ref{tab:mag_ini_colors_linestyles}; for each line in that table (colour), the further right the column is (the higher $B_0$ or $B_\textrm{opp}$, the changing numbers in the run names), the higher the inclination.}
    \label{fig:mag_ini}
\end{figure*}
\begin{table*}[h]
    \centering
    \caption{Colours and line styles for simulations given in bottom right panel of Fig.~\ref{fig:mag_ini}, showing the inclination in the initial field.}
    \label{tab:mag_ini_colors_linestyles}
    \setlength{\tabcolsep}{2.9pt}%
    \begin{tabular}{l|llllll|l}
\hline\hline%
colour \iflatexml$\Big\backslash$\else\begin{picture}(7,8)\put(0,8){\line(1,-2){5.7}}\end{picture}\fi line style & 
\textemdash & $\cdots$ & - - & \raisebox{0.08ex}{-}\,$\cdot$\,\raisebox{0.08ex}{-}\,$\cdot$ & $-\cdots$ & $-$\ $-$ & 
comment\\\hline\vphantom{$1^1$}%
light red & \hphantom{1}20\_opp000o & \hphantom{1}40\_opp000o & \hphantom{1}80\_opp000o &&&& $B_0<160\,$kG\\
dark red  & 160\_opp000? & 160\_opp050  & 160\_opp100  & 160\_opp150  & 160\_opp200  & 160\_opp300\\
light blue &              &              & 160\_opp100w & 160\_opp150w & 160\_opp200w &&  $w=98.304\,$Mm\\
dark blue  &              & 160\_opp050c & 160\_opp100c & 160\_opp150c & 160\_opp200c & 160\_opp300c & $F_\textrm{Gauss}>10^{22}$\,Mx\hspace*{-2.1pt}\\\hline
    \end{tabular}
\end{table*}
\bigskip
\newlength\myMPwidth\iflatexml\setlength{\myMPwidth}{\textwidth}\else\setlength{\myMPwidth}{9cm}\fi
\noindent\begin{minipage}[t]{\myMPwidth}%
Figure~\ref{fig:mag_ini} shows the azimuthal averages of the initial magnetic field at the average height of the $\tau=1$ iso-surface in the quiet Sun.
The top two panels ($B_\mathrm{z}+B_\textrm{opp}$ and $B_\mathrm{r}$), show the initial conditions for all our simulations (except `80\_opp000o' and `alpha*'), because introducing a $B_\textrm{opp}\ne0$ only offsets $B_\mathrm{z}$ uniformly, and does not affect $B_\mathrm{r}$.
In the bottom right panel, all simulations (except `alpha*') are plotted. Colours and line-styles are described in Table ~\ref{tab:mag_ini_colors_linestyles}, where the `?' in `160\_opp000?' is a wildcard, standing for any of `o', `ho', `h', `', or `b', since these simulations share the same initial conditions.\sethyphenpenalty{1000}\par
\end{minipage}\iflatexml\par\noindent\else\hspace{4mm}\fi\begin{minipage}[t]{\myMPwidth}%
\hspace{15pt}While this figure shows many details, primarily the following aspects are pertinent: 
\begin{enumerate}
    \item[1)] $F_\textrm{Gauss}>10^{22}$\,Mx leads to strong $B_\mathrm{z}$ on the spot axis. 
    \item[2)] $w=98.304\,$Mm leads to the field being distributed over a wider box, resulting in a weaker $B_\mathrm{z}$ far from the spot axis and more extended $B_\mathrm{r}$.
    \item[3)] The inclinations in the initial conditions show a wide range of behaviour, in particular outside the spot.
\end{enumerate}
\end{minipage}

\end{appendix}
\end{document}